\documentclass[3p, twocolumn]{elsarticle}
\usepackage[english]{babel}
\usepackage[utf8]{inputenc}
\usepackage{graphicx,epstopdf,color}
\usepackage{xspace}
\usepackage{float}
\usepackage{hyperref}

\begin{document}

\begin{frontmatter}
\title{Charge Ordering and Magnetic Exchange in the Ladder-Type Compound NH$_4$V$_2$O$_5$}

\author{Dm. M.~Korotin}
\ead{dmitry@korotin.name}

\address{M.N. Mikheev Institute of Metal Physics, S.Kovalevskoy St. 18, 620137 Yekaterinburg, Russia}

\address{Skolkovo Institute of Science and Technology, 30 Bolshoy Boulevard, bld. 1, Moscow 121205, Russia}

\date{\today}

\begin{abstract}
The low-temperature electronic and magnetic properties of NH$_4$V$_2$O$_5$, an isoelectronic analog of the spin-ladder compound $\alpha'$-NaV$_2$O$_5$, are investigated using DFT+$U$ calculations. Two charge-ordering patterns -- zigzag and linear chains of V$^{4+}$/V$^{5+}$ ions -- are considered. The zigzag configuration is found to be energetically preferred and exhibits insulating behavior with a band gap of 1.7~eV. In this state, magnetic V$^{4+}$ (d$^1$) ions form antiferromagnetically coupled spin chains. Calculated exchange interactions reveal strong diagonal (intrarung) and interladder couplings, indicating a complex spin-ladder network. These results suggest that NH$_4$V$_2$O$_5$ retains essential spin-ladder characteristics while displaying new structural and magnetic features arising from the larger size and non-spherical geometry of the NH$_4^+$ ion compared to Na$^+$.
\end{abstract}

\begin{keyword}
Spin-ladder; Charge ordering; DFT; Electronic properties; Exchange interaction
\end{keyword}

\end{frontmatter}

\section{Introduction}

Low-dimensional quantum spin systems exhibit complex interplay among spin, charge, and lattice degrees of freedom. This coupling gives rise to emergent phenomena such as charge ordering, spin-gap formation, and spin-Peierls transitions. Among these systems, $\alpha'$-NaV$_2$O$_5$ serves as a prototypical inorganic spin-ladder compound, where the vanadium-oxygen network forms quasi-one-dimensional two-leg ladders~\cite{Isobe1996,Smolinski1998,Ludecke1999}.

\begin{figure}[h]
    \centering
    \includegraphics[width=\linewidth]{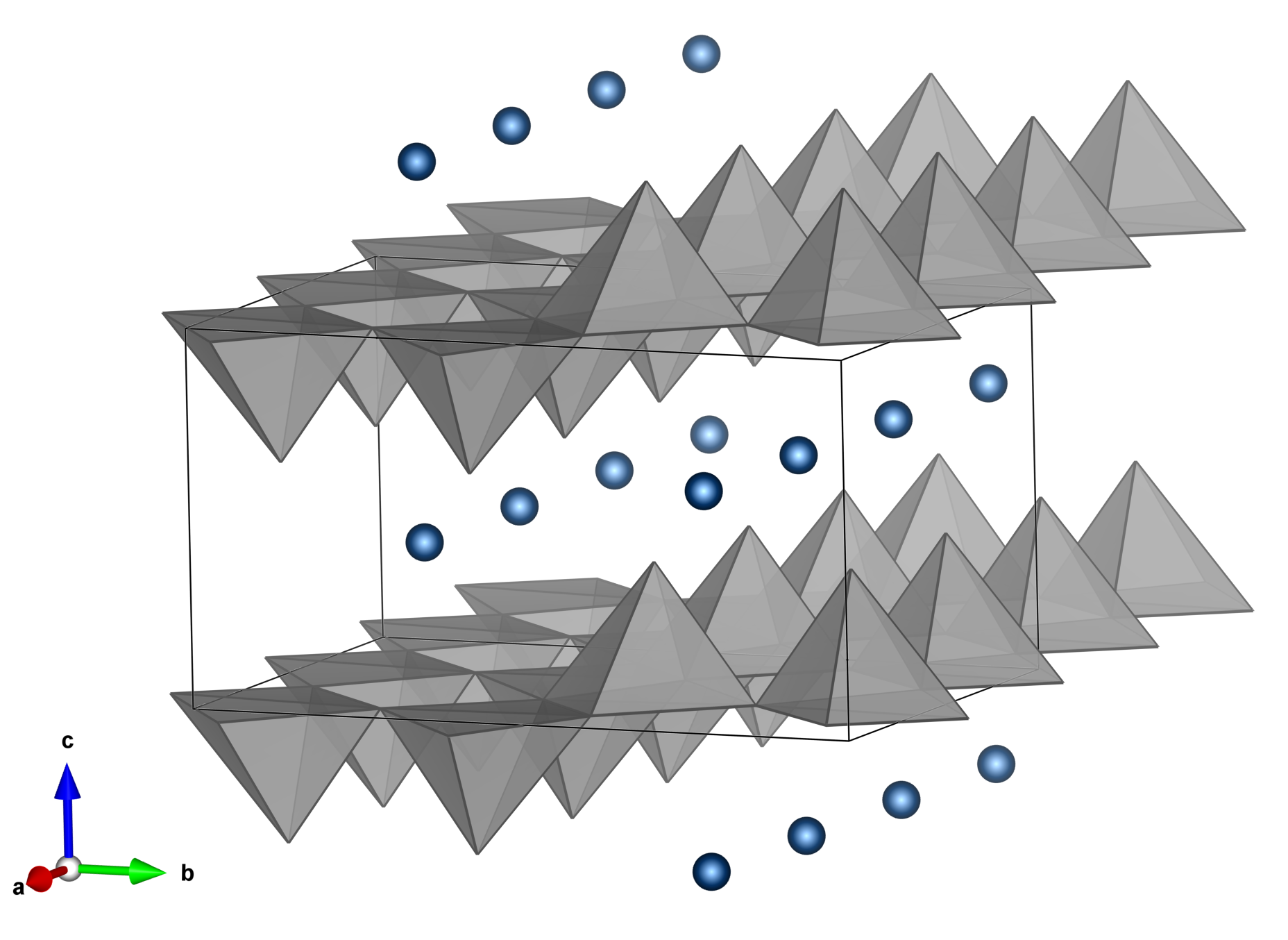}
    \caption{
    Representative two-leg ladder crystal structure of XV$_2$O$_5$ compounds in the $Pmmn$ space group, where X represents Na, Ca, or NH$_4$. Vanadium atoms are situated within oxygen pyramids (gray), while X ions (blue spheres) are positioned between the V-O layers.}
    \label{fig:proto_struc}
\end{figure}

The parent compound in the vanadate family, V$_2$O$_5$, consists of corner- and edge-shared VO$_5$ pyramids with vanadium ions in the formal +5 oxidation state, forming nonmagnetic $S = 0$ configurations (Figure~\ref{fig:proto_struc}). Upon intercalation of alkali or alkaline earth cations, one or two electrons per formula unit are donated to the vanadium-oxygen layers. This additional electron density populates vanadium 3$d$ states and modifies both electronic and magnetic characteristics. The resulting network of antiferromagnetically coupled $S = 1/2$ spins appears on a ladder structure, with exchange interactions determined by the occupation of specific vanadium sites and V-O-V bond angles and distances~\cite{Ueda1998a,Korotin1999,Konstantinovic2000,Takeo1999}. The effects of partial occupation of vanadium sites on the ladder can be also distinguished through vacancies in the oxygen sublattice~\cite{Cezar2014,Xiao2009} or via injection of excess electrons into vanadium pentoxide thin films~\cite{Bhandari2015,Ritika2025}.

A recently synthesized V$_2$O$_5$-based compound, NH$_4$V$_2$O$_5$, which is isoelectronic to $\alpha'$-NaV$_2$O$_5$, has attracted considerable interest in the field of energy storage materials~\cite{Li2020,wang2023,Tan2024}. From both structural and electrochemical perspectives, ammonium ions intercalated between the V–O layers increase the interlayer spacing, alter energy barriers for Li$^+$ and Zn$^{2+}$ intercalation, and significantly enhance battery performance. Notably, each ammonium ion donates an electron to the vanadium $d$-subshell, suggesting that the low-temperature electronic and magnetic properties of NH$_4$V$_2$O$_5$ may resemble those of $\alpha'$-NaV$_2$O$_5$.

However, the Na$^+$ ion has a significantly smaller ionic radius than NH$_4^+$ (116pm vs 168pm)~\cite{Sidey2016}, which can lead to greater distortion of the host vanadate crystal structure and, as a result, modify electron transfer integrals and magnetic exchange interactions between vanadium sites. Since research in energy storage materials has primarily focused on room-temperature properties, experimental data at low temperatures remain limited. In this context, \textit{ab initio} calculations serve as a crucial tool for predicting the magnetic and charge-ordering behavior of NH$_4$V$_2$O$_5$.

This study employs density functional theory calculations to investigate the electronic structure, magnetic exchange interactions, and structural features of NH$_4$V$_2$O$_5$. By comparing these properties with those of $\alpha'$-NaV$_2$O$_5$, the analysis explores whether spin-ladder behavior and charge-ordering tendencies are retained following ammonium substitution. 

\section{Methods}

The electronic structure calculations were performed using the \textsc{Quantum ESPRESSO} package~\cite{Giannozzi2009,giannozzi2020}. Pseudopotentials were taken from the Standard Solid-State Pseudopotentials (SSSP) library, using the PBEsol Precision set (v1.3.0)\cite{prandini2018}. The kinetic energy cutoffs were set to 80Ry for the plane-wave basis and 600~Ry for the charge density expansion. Brillouin zone integration was carried out on a $6 \times 6 \times 8$ Monkhorst-Pack $k$-point mesh within the irreducible part of the Brillouin zone.

Long-range dispersion interactions were included via Grimme’s D3 correction scheme with three-body terms~\cite{grimme2010}, ensuring an accurate description of interlayer van der Waals forces. Structural optimizations were performed until the total energy change was below $10^{-5}$~Ry, residual atomic forces were less than $10^{-3}$Ry/Bohr, and the pressure on the simulation cell was reduced below 0.2kbar. All simulations were carried out on a supercell containing four formula units, obtained by doubling the primitive cell along the [100] direction.

To account for the localized nature of the V~$3d$ electrons, the DFT+$U$ approach was employed in the Dudarev formulation~\cite{DudarevLDA+U}, with an effective Hubbard parameter $U_{\mathrm{eff}} = 3.2$eV. This value is consistent with previous studies on vanadium oxides\cite{wang2023,wu2019,Anisimova2025}, where it was shown to reliably describe charge localization and magnetic ordering in mixed-valence systems.

The magnetic exchange interaction parameters $J_{ij}$ of the Heisenberg model were calculated using the Green’s function formalism, as implemented in the \textsc{EXCHANGES} code~\cite{exchanges}. Since the Heisenberg model is defined for localized spin moments, a mapping from DFT results requires site-centered orbitals. To achieve this, projected Wannier functions with the symmetry of V~$d$ and O~$p$ atomic orbitals were constructed~\cite{hamilt}.

Noncovalent interaction (NCI) plots were generated from total charge density data using the Critic2 package~\cite{otero-de-la-roza2014}.

\section{Results}

An electron from the intercalated ammonium ion in the V$_2$O$_5$ structure occupies the originally empty $d$-shell of a V atom. In the pyramidal coordination environment, the lowest-energy orbital is $d_{xy}$. Since there is only one additional electron per two V atoms in NH$_4$V$_2$O$_5$, the question of the resulting charge distribution between vanadium sites remains open. The isoelectronic compound $\alpha'$-NaV$_2$O$_5$ exhibits all V atoms in the same formal V$^{4.5+}$ ($d^{0.5}$) valence configuration at room temperature~\cite{Ohama1999,Smolinski1998}. However, there have been debates regarding its low-temperature phase (below $T_c \approx 34$~K): whether it exhibits V$^{5+}$/V$^{4+}$ charge ordering  with zigzag~\cite{Smirnov1999,Mostovoy1999a} or linear chains~\cite{Carpy1972} of V ions in $d^1$ electronic configuration.

\begin{figure}[h]
    \centering
    \includegraphics[width=\linewidth]{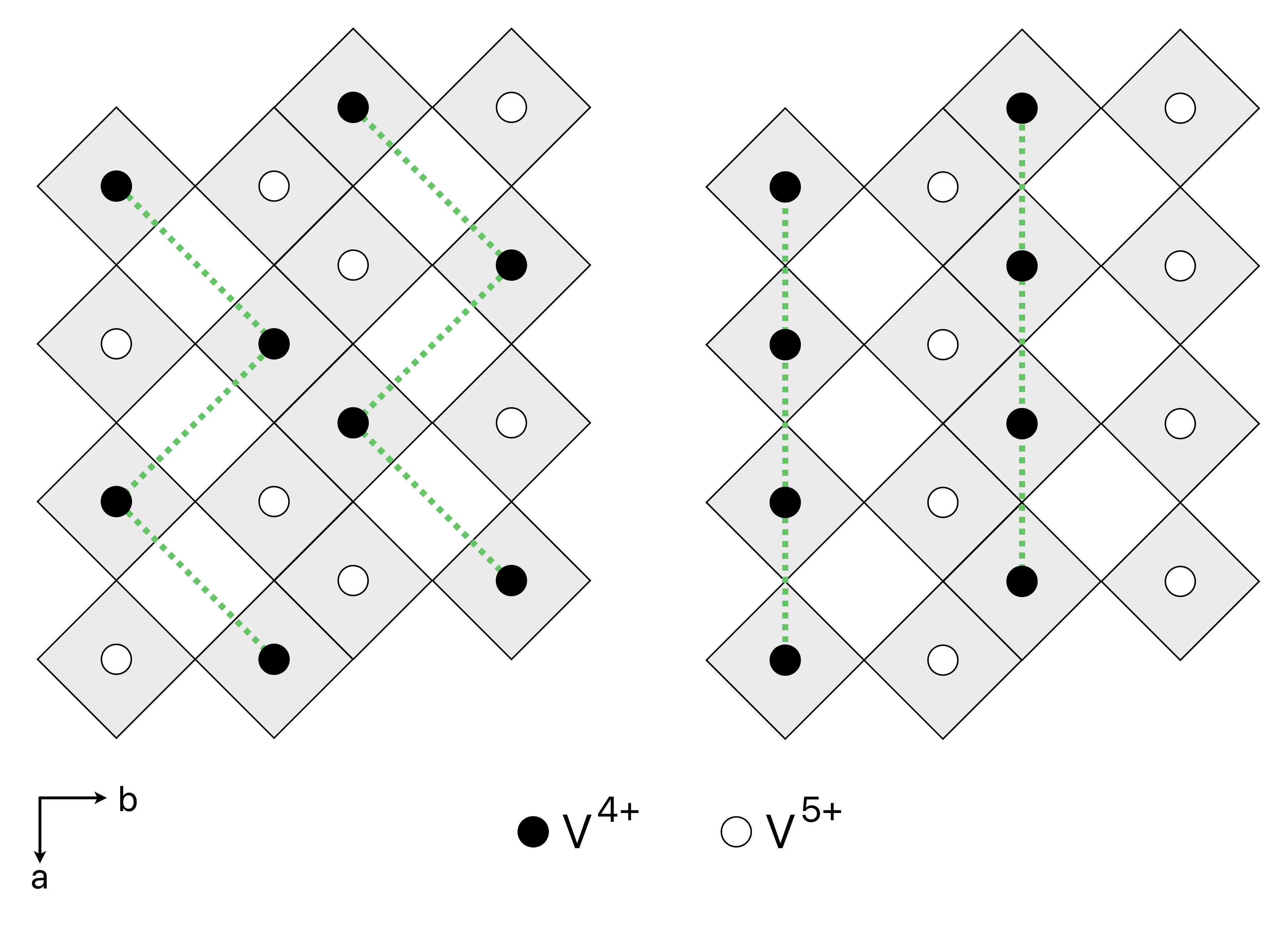}
    \caption{
    The two charge orderings of V$^{5+}$/V$^{4+}$ ions considered in the $ab$-plane of NH$_4$V$_2$O$_5$: zigzag (left) and linear chain (right) arrangements. Gray squares represent V-O pyramid base planes, while filled and empty circles indicate V$^{4+}$ and V$^{5+}$ ions, respectively. Green dashed lines highlight the zigzag and linear chain structures to guide the eye.}
    \label{fig:orderings}
\end{figure}

Building on these quarter-century-old discussions and considering that the ammonium ion is (i) 30\% larger than sodium and (ii) non-spherical, which could reduce symmetry and distort the local structure, we compared two possible charge orderings in NH$_4$V$_2$O$_5$ shown in Figure~\ref{fig:orderings}.

In the absence of structural data for NH$_4$V$_2$O$_5$, we based our approach on results from our own ab initio relaxation calculations.
We started from the crystal structure of pristine vanadium pentoxide~\cite{villarsa}, then used Voronoi decomposition to identify five inequivalent interstitial sites for ammonium ion insertion between the V-O layers. The initial positions of NH$_4^+$ ions were chosen to maximize the distance from neighboring vanadium and oxygen atoms, thereby minimizing local strain. Structural relaxations were performed within the DFT framework, allowing both atomic positions and lattice parameters to relax. We then chose the structure with the lowest enthalpy.

Since DFT underestimates correlation effects in the vanadium $d$-shell, the resulting structure contains all equivalent V atoms with a formal valency of V$^{4.5+}$ -- there is no evidence of charge ordering.

The lattice parameters changed compared to pristine V$_2$O$_5$: $a$ increased by 2.5\%, from 3.56~\AA\ to 3.65~\AA; $b$ decreased by 3.4\% (11.52~\AA\ $\to$ 11.10~\AA); the $c$ lattice vector (or interlayer distance) grew more significantly from 4.37 to 5.97~\AA, with the main reason being NH$_4^+$ ion insertion. The V-O pyramids in the obtained structure have a volume of 4.87~\AA$^3$ compared to 4.48~\AA$^3$ in the pristine structure -- this indicates that every vanadium atom receives an equal additional amount of charge in DFT.

To improve the description of charge localization and to account for electronic correlation effects in the partially filled $d$-subshell of vanadium atoms, we performed a full atomic relaxation of the previously obtained crystal structure using the DFT+$U$ approach. This method favors integer occupation of $d$-orbitals, thus promoting electron localization. The localization of electrons on two of the four vanadium sites in the unit cell and the appearance of the mixed valence state were expected.

Magnetic systems exhibit complex potential energy landscapes with numerous local minima, each corresponding to distinct magnetic orderings. Due to the strong coupling between electronic and lattice degrees of freedom, atomic relaxation calculations frequently remain trapped near the initial configuration, making it challenging to escape local minima and explore the full configuration space.

To address this limitation, we employed constrained DFT+U calculations. To model the zigzag and linear chain types of charge ordering shown in Figure~\ref{fig:orderings}, starting from an unordered crystal structure, we constrained the magnetic moment (and correspondingly the number of $d$-electrons) for half of the V ions. Following relaxation of all atomic coordinates and lattice vectors, the crystal structure adopts the desired charge ordering. In the final step, all constraints are removed, and full relaxation continues until reaching a local minimum in configuration space. Comparison of the resulting enthalpies allows determination of the most stable electronic/magnetic/crystal structure.

\begin{figure}[h]
    \centering
    \includegraphics[width=\linewidth]{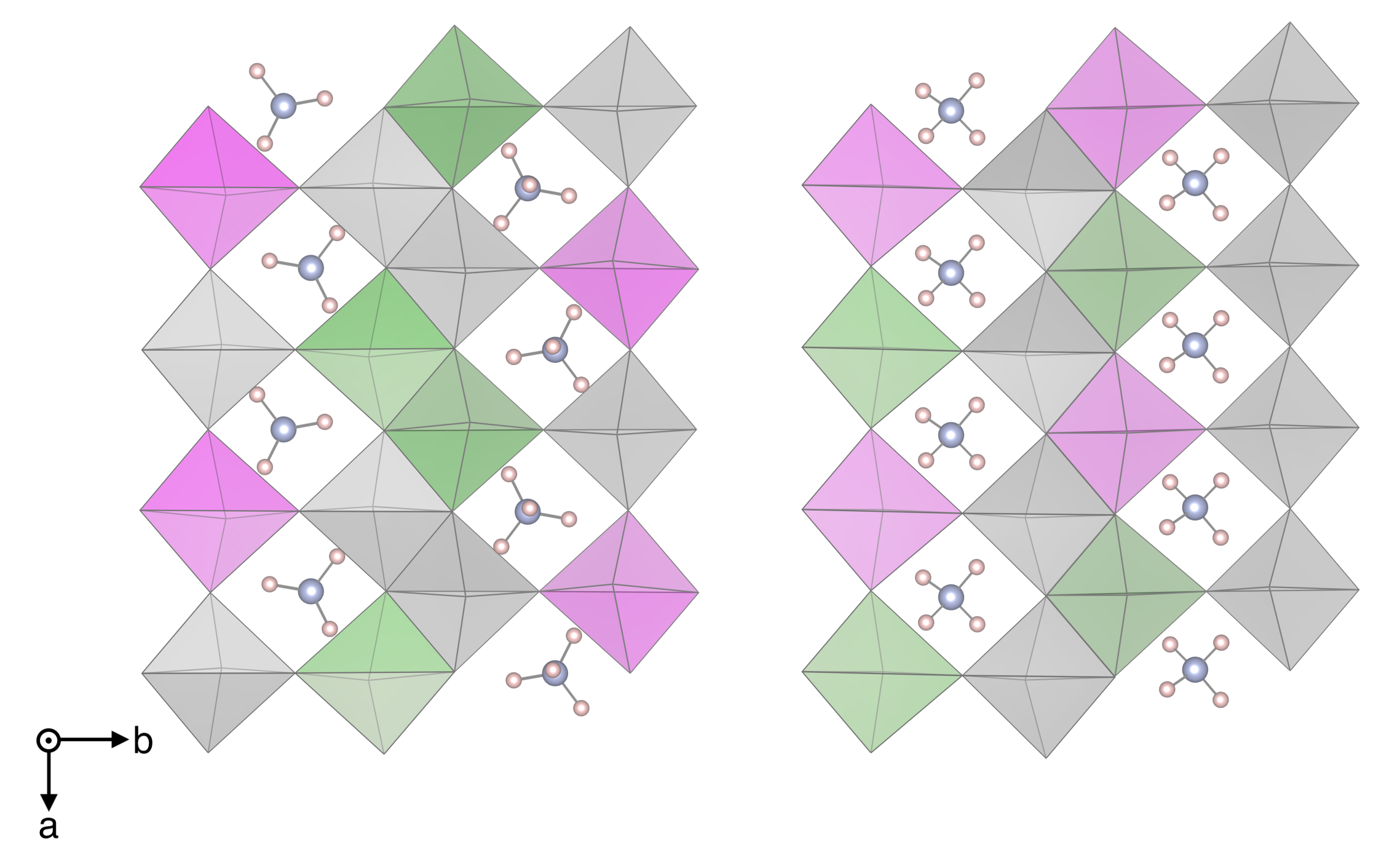}
    \caption{Optimized crystal structures showing two competing charge and magnetic ordering patterns in NH$_4$V$_2$O$_5$: zigzag ordering (left) and linear chain ordering (right). Magenta and green oxygen pyramids contain magnetic V$^{4+}$ ions with opposite spin orientations, while gray pyramids denote nonmagnetic V$^{5+}$ ions. Small blue and salmon spheres represent intercalated NH$_4^+$ ions positioned between the vanadium-oxygen layers.}
    \label{fig:2orderings}
\end{figure}

The optimized crystal structures for both charge ordering scenarios are presented in Figure~\ref{fig:2orderings}. In both configurations, the charge ordering results in a 1:1 ratio of V$^{4+}$:V$^{5+}$ sites, with the magnetic V$^{4+}$ ions (d$^1$ configuration) exhibiting antiferromagnetic coupling within the ladder. The zigzag ordering (left panel) shows alternating spin-up and spin-down V$^{4+}$ sites arranged in diagonal chains across the $ab$-plane, while the linear chain ordering (right panel) displays these magnetic sites in more regular linear arrangements. The nonmagnetic V$^{5+}$ ions (d$^0$ configuration) occupy the remaining vanadium sites. Both structures preserve the characteristic layered framework with NH$_4^+$ ions intercalated between the vanadium-oxygen layers. 
Notably, the NH$_4^+$ ion arrangement differs significantly between the two phases: in the zigzag ordering, the ammonium ions adopt a zigzag-like pattern along the $a$ lattice vector, whereas in the linear chain ordering, they align in straight lines.
The structural optimization reveals that the charge ordering affects the local coordination environment, with distinct V-O bond lengths for the different oxidation states, while the antiferromagnetic coupling between neighboring V$^{4+}$ sites stabilizes the magnetic ground state in both ordering patterns.

\begin{table*}[h]
\centering
\caption{Comparison of structural and electronic parameters between zigzag and linear chain configurations of V$^{4+}$ ions in NH$_4$V$_2$O$_5$, calculated using spin-polarized DFT+U.}

\label{tab:parameters}
\begin{tabular}{l|c|c}
\hline \hline

\textbf{Parameter} & \textbf{Zigzag} & \textbf{Linear chain} \\ \hline
Final enthalpy, meV& 0 & +330 \\ \hline
$a$, \AA & 7.37  & 7.40 \\ \hline
$b$, \AA & 11.14 & 11.13\\ \hline
$c$, \AA & 5.96  & 6.02 \\ \hline
Cell volume, \AA$^3$ & 489.75 & 495.62 \\ \hline
V-O pyramids volume, \AA$^3$ &  4.81 \& 5.21 &  4.82 \& 5.17 \\ 
\hline
Charge of V ions & 8.44 \& 8.54 & 8.42 \& 8.52 \\ \hline
V-$d$ occupation (Lödwin) & 3.25 \& 3.28 & 3.25 \& 3.30 \\ 

\hline \hline
\end{tabular}
\end{table*}

Data comparing the two ordered structures are presented in Table~\ref{tab:parameters}. The zigzag ordered phase is more energetically favorable than the linear chain phase, suggested in an earlier paper~\cite{Anisimova2025}, with an enthalpy gain of 330~meV per unit cell containing 4 formula units.

\begin{figure}[h]
    \centering
    \includegraphics[width=\linewidth]{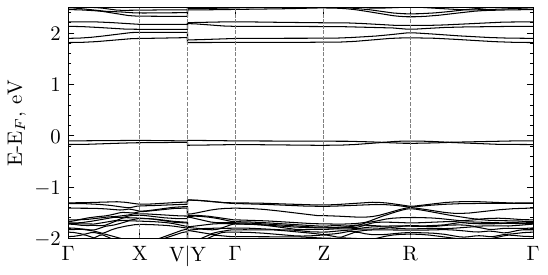}
    \caption{Band structure for the spin-up channel of NH$_4^+$-V$_2$O$_5$. The band structure for the spin-down channel is the same, since the compound is antiferromagnetic.}
    \label{fig:bands}
\end{figure}

The formation of two distinct types of V ions is evident from the difference in V--O pyramid volumes: pyramids containing V$^{5+}$ ions contract from 4.87 to 4.81~\AA$^3$ compared to the unordered DFT structure, while those containing V$^{4+}$ ions expand from 4.87 to 5.21~\AA$^3$. However, the splitting of V ions into two oxidation states is not clearly reflected in either the calculated ionic charges (determined by integrating the charge density within spheres around each ion) or the L{\"o}wdin population analysis of the V-$d$ subshells.

\begin{figure}[h]
    \centerline{\includegraphics[width=\linewidth]{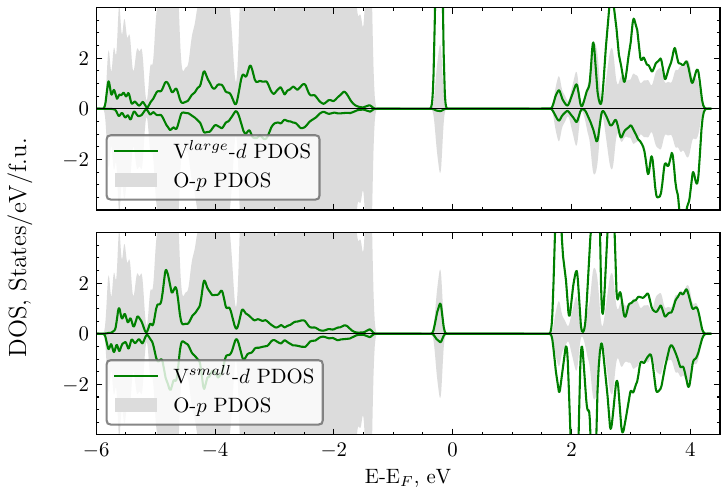}}
\caption{Spin-resolved partial densities of states (PDOS) for vanadium ions occupying large (V$^{{large}}$) and small (V$^{{small}}$) pyramids in the structure of NH$_4$V$_2$O$_5$. The PDOS for O-$2p$ states is obtained by summing contributions from all oxygen atoms in the simulation cell. The upper part of each panel corresponds to the spin-up channel and the lower part to the spin-down channel.}
    \label{fig:pdoses}
\end{figure}

The partial densities of states for V ions located in large and small pyramids shown in Figure~\ref{fig:pdoses} clearly reveal the difference in electronic configuration between the two types of V sites.

In the zigzag ordered structure, NH$_4$V$_2$O$_5$ exhibits insulating behavior with a direct energy gap of approximately 1.7~eV as shown in Figure~\ref{fig:bands}. The sharp peak in the DOS below the Fermi level (and the corresponding ultranarrow energy band) is predominantly formed by spin-up $d$ states of V ions located in large pyramids. This confirms that these sites have a $d^1$ configuration, corresponding to the V$^{4+}$ oxidation state. The vanadium ions inside the smaller pyramids display an almost spin-unpolarized partial DOS, with only a tiny peak below the Fermi level, possibly indicating hybridization of $d$-states between V$^{{large}}$ and V$^{{small}}$ ions. Thus, these ions can be confidently assigned to the V$^{5+}$ ($d^0$) state.
Electronic states below $-1.4$~eV (relative to the Fermi level) are predominantly composed of oxygen $p$ states with minor contributions from V-$d$ states. Conversely, electronic states above the Fermi level consist primarily of V-$d$ states with an admixture of O-$p$ states. The partial DOSes indicate strong hybridization between the metal and ligand states and even the peak at $-0.5$~eV has a significant contribution of oxygen states.

\begin{figure}[h]
    \centering
    \includegraphics[width=0.95\linewidth]{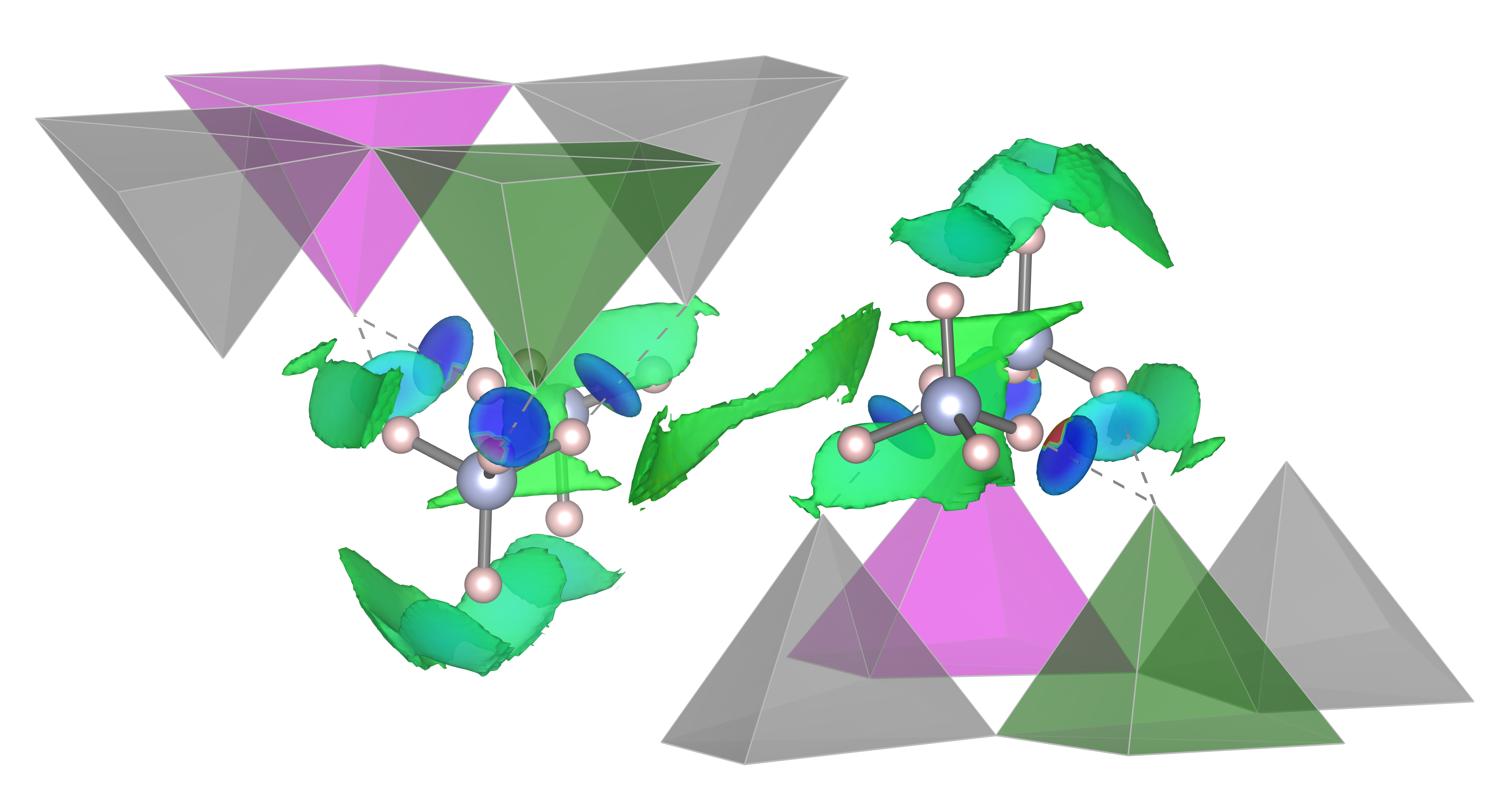}
    \caption{Noncovalent interaction (NCI) regions plot for the zigzag-ordered phase of NH$_4$V$_2$O$_5$, showing only interactions between the NH$_4^+$ cations and their nearest oxygen atoms. Magenta and green VO$_5$ pyramids highlight magnetic V$^{4+}$ ions with opposite spin orientations, while gray pyramids denote nonmagnetic V$^{5+}$ ions. Small blue (N) and salmon (H) spheres represent the intercalated NH$_4^+$ cations positioned between the vanadium-oxygen layers. The isosurface represents the reduced density gradient $s(\rho)$ at $s = 0.5$, color-mapped according to the $\mathrm{sign}(\lambda_2)\rho$ criterion. Blue regions correspond to attractive interactions dominated by N-H$\cdots$O hydrogen bonding, while green lobes denote weak dispersive (van der Waals) interactions. The directional N–H$\cdots$O hydrogen bonds shown with dashed lines.}
    \label{fig:nci_zigzag}
\end{figure}

The noncovalent interaction (NCI) plot presented in Figure~\ref{fig:nci_zigzag} shows the interactions between intercalated NH$_4^+$ cations and nearby framework oxygen atoms. The NCI isosurface, colored using the sign($\lambda_2$)$\rho$ criterion~\cite{Johnson2010}, shows localized blue regions along N–H,$\cdots$,O directions, characteristic of attractive hydrogen bonding, and green lobes indicative of weak dispersive interactions. The absence of extended red features suggests that steric repulsion is minor and confined to small regions. Quantitatively, each ammonium ion forms three short H$\cdots$O contacts of 1.721, 1.785, and 1.994~\AA, while the fourth hydrogen atom engages in a longer, weaker interaction of 2.237~\AA. Given that typical hydrogen bonds in oxide matrices fall within $\sim$1.6–2.0~\AA, these results confirm that each NH$_4^+$ establishes three (of possible four) directional N–H$\cdots$O hydrogen bonds together with an additional, weaker contact.
These hydrogen bonds involve oxygen atoms coordinated to both expanded VO$_5$ pyramids (V$^{4+}$-like sites) and contracted VO$_5$ pyramids (V$^{5+}$-like sites). Within the accuracy of the NCI analysis and the computed H$\cdots$O distances, however, no systematic difference is observed between interactions involving V$^{4+}$- and V$^{5+}$-associated oxygens.

\begin{figure}[h]
    \centering
    \includegraphics[width=\linewidth]{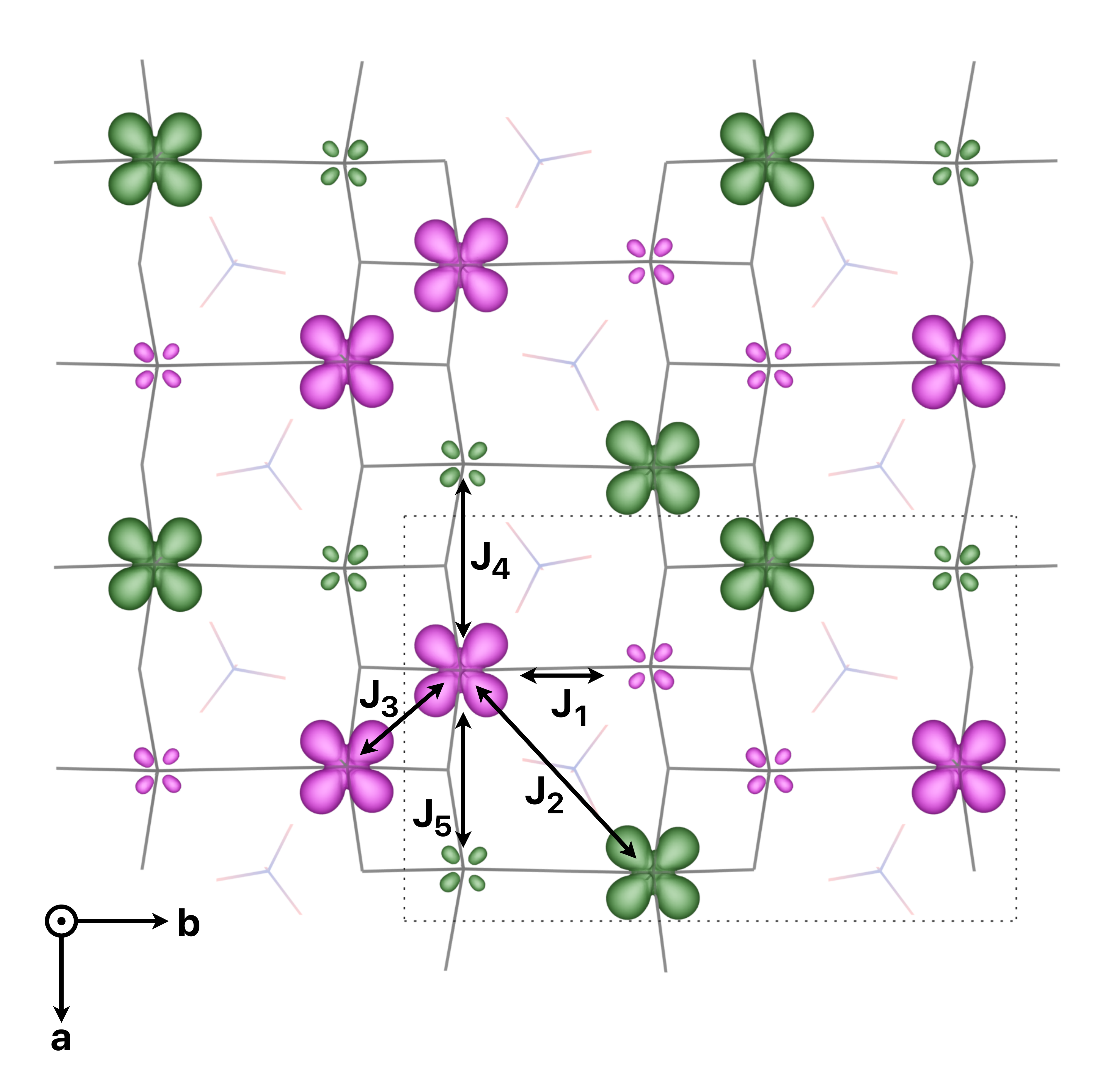}
    \caption{Calculated magnetization density $\rho^{\uparrow} - \rho^{\downarrow}$ (green for positive and magenta for negative values) for NH$_4$V$_2$O$_5$. The isosurface corresponds to 5\% of the maximum polarization value. Black lines represent vanadium-oxygen bonds, and the directions of magnetic exchange interactions $J_1$--$J_5$ are indicated by black arrows. The lattice cell is shown with a black dotted line.}
    \label{fig:spinrho}
\end{figure}

The spatial arrangement of antiferromagnetic zigzag chains is revealed through the magnetization density isosurface ($\rho^{\uparrow} - \rho^{\downarrow}$) presented in Figure~\ref{fig:spinrho}. The unpaired electrons occupy the $d_{xy}$ orbitals of the magnetic V$^{4+}$ sites, while the formal V$^{5+}$ ions exhibit minimal magnetization with values significantly smaller than those of the V$^{4+}$ ions.

Using the Green's function approach, we calculated the exchange interaction parameters $J_{ij}$ for the Heisenberg model with the Hamiltonian:
\[\label{LP}
H = -\sum_{\langle ij\rangle} J_{ij} {\bf m}_i {\bf m}_j, 
\]
where ${\bf m}_i$ represents the magnetic moment of the $i$th site, and the summation runs over each ion pair. The calculated intraladder ($J_1$, $J_2$, $J_4$, $J_5$) and interladder ($J_3$) exchange parameters are presented in Table~\ref{tab:exchanges}, with corresponding directions shown in Figure~\ref{fig:spinrho}.

\begin{table}[h]
\centering
\caption{Calculated exchange interaction parameters (meV) between V ions in NH$_4$V$_2$O$_5$, corresponding magnetic moments of the coupled ions ($m_i, m_j$), and interatomic distances.}
\label{tab:exchanges}
\begin{tabular}{l|c|c|c}
\hline \hline

& \textbf{$J_{ij}$ (meV)} & $| m_i | , | m_j| $ ($\mu_B$) & \textbf{distance (\AA)} \\ \hline
$J_1$ &  9.9 & 1.06, 0.08 & 3.45 \\
$J_2$ & -3.2 & 1.06, 1.06 & 5.08 \\
$J_3$ &  3.3 & 1.06, 1.06 & 3.06 \\
$J_4$ &  0.1 & 0.08, 0.08 & 3.76 \\
$J_5$ &  0.3 & 0.08, 0.08 & 3.62 \\
\hline \hline
\end{tabular}
\end{table}

The largest exchange interaction parameter is $J_1 = 9.9$~meV, corresponding to coupling across the ladder rungs. However, since this interaction occurs between large ($m_i = 1.06~\mu_B$) and small ($m_j = 0.08~\mu_B$) magnetic moments, its contribution to the total magnetic energy is substantially smaller than that from the diagonal interaction between neighboring rungs, $J_2 = -3.2$~meV (negative sign indicating antiferromagnetic exchange). Notably, the interladder exchange interaction $J_3$ has a comparable magnitude to $J_2$, suggesting significant magnetic coupling between adjacent ladders. The calculated exchange interaction parameter between the layers (along the [001] direction) is 0.003~meV, therefore this interaction is negligible.

These results demonstrate that the magnetic structure of NH$_4$V$_2$O$_5$ is governed by a combination of strong rung interactions and substantial diagonal and interladder couplings. The presence of considerable interladder exchange may lead to more complex magnetic ordering patterns than observed in the parent $\alpha'$-NaV$_2$O$_5$ compound, where interladder interactions are typically weaker. 

\section{Conclusion}
In this work, the low-temperature electronic and magnetic properties of NH$_4$V$_2$O$_5$ were investigated using first-principles calculations. Motivated by its structural similarity and isoelectronic character to the prototypical spin-ladder compound $\alpha'$-NaV$_2$O$_5$, this study focused on the emergence of charge ordering, magnetic exchange interactions, and the stability of spin-ladder motifs upon ammonium intercalation.

Starting from an ab initio relaxed structure, DFT+$U$ calculations with constrained electron localization were employed to explore two competing charge-ordered states: a zigzag configuration and a linear chain configuration of V$^{4+}$/V$^{5+}$ ions. The calculations reveal that the zigzag configuration is energetically more favorable by 330~meV per unit cell, stabilized by antiferromagnetic ordering among the magnetic V$^{4+}$ sites. The electronic structure analysis confirms the formation of an energy gap of $\sim$1.7~eV in the zigzag phase, with magnetic moments localized primarily on vanadium sites in expanded VO$_5$ pyramids, consistent with a $d^1$ ($S = 1/2$) configuration.

The computed exchange interactions show a dominant antiferromagnetic diagonal coupling ($J_2$) within the ladder, alongside non-negligible interladder coupling ($J_3$), indicating a more complex magnetic network than in the isoelectronic NaV$_2$O$_5$ compound. Thus, while the basic ladder-like spin topology is preserved, the larger ammonium ion, through its effect on interlayer spacing and local symmetry, introduces substantial modifications to both the electronic structure and magnetic interactions.

The role of the NH$_4^+$ cation extends beyond its larger size and non-spherical geometry to include its ability to form directional N--H$\cdots$O hydrogen bonds. Although these bonds are relatively weak, they could subtly polarize the bridging oxygen atoms and thereby modulate the effective V--O $d$--$p$ overlap and O~2$p$ levels. Such subtle changes, combined with structural distortions induced by NH$_4^+$, may provide a plausible explanation for the differences in exchange couplings $J$ and for the altered magnetic exchange network observed in comparison with NaV$_2$O$_5$.

\section*{Acknowledgments}
Results were obtained with the support of the Russian Science Foundation (Project No. 24-22-00349).

\section*{Data Availability}
All calculations were performed using codes available publicly. To reproduce the results of the work, one should use the calculation parameters presented in the Methods section.

\section*{Declaration of generative AI and AI-assisted technologies in the writing process}
During the preparation of this work the author used Claude 3.5 Sonnet tool to assist with grammar checking and language refinement. After using this tool, the author reviewed and edited the content as needed and takes full responsibility for the content of the published article.

\bibliographystyle{elsarticle-num}
\bibliography{main}

\end{document}